\documentclass[10pt,a4paper,english,conference]{IEEEtran}
\usepackage[T1]{fontenc}
\usepackage[latin9]{inputenc}
\usepackage{color}
\usepackage{babel}
\usepackage{verbatim}
\usepackage{float}
\usepackage{mathrsfs}
\usepackage{amsthm}
\usepackage{amsmath}
\usepackage{amssymb}
\usepackage{graphicx}
\usepackage{setspace}
\usepackage{esint}
\usepackage{epstopdf}
\usepackage{subfigure}
\usepackage{balance}

\makeatletter

\pdfpageheight\paperheight
\pdfpagewidth\paperwidth

\floatstyle{ruled}
\newfloat{algorithm}{tbp}{loa}
\providecommand{\algorithmname}{Algorithm}
\floatname{algorithm}{\protect\algorithmname}

\theoremstyle{plain}
\newtheorem{thm}{\protect\theoremname}
\theoremstyle{remark}

\theoremstyle{plain}
\newtheorem{lem}[thm]{\protect\lemmaname}

\makeatother

\providecommand{\remarkname}{Remark}
\providecommand{\lemmaname}{Lemma}
\providecommand{\theoremname}{Theorem}

\begin{document}

\title{Will the Area Spectral Efficiency Monotonically Grow as Small Cells Go Dense?
}

\author{\noindent {\normalsize Ming Ding, \emph{National ICT Australia (NICTA), Australia} \{\texttt{Ming.Ding@nicta.com.au}\}}\\
{\normalsize David L$\acute{\textrm{o}}$pez-P$\acute{\textrm{e}}$rez,
\emph{Bell Labs Alcatel-Lucent, Ireland} \{\texttt{dr.david.lopez@ieee.org}\}}\\
{\normalsize Guoqiang Mao, \emph{The University of Technology Sydney and NICTA, Australia} \{\texttt{g.mao@ieee.org}\}}\\
{\normalsize Peng Wang, \emph{The University of Sydney and NICTA, Australia} \{\texttt{thomaspeng.wang@sydney.edu.au}\}}\\
{\normalsize Zihuai Lin, \emph{The University of Sydney, Australia} \{\texttt{zihuai.lin@sydney.edu.au}\}}\\
}



\maketitle

\begin{abstract}
In this paper, we introduce a sophisticated path loss model into the stochastic geometry analysis incorporating both line-of-sight (LoS) and non-line-of-sight (NLoS) transmissions
to study their performance impact in small cell networks (SCNs).
Analytical results are obtained on the coverage probability and the area spectral efficiency (ASE) assuming both a general path loss model and a special case of path loss model recommended by the 3rd Generation Partnership Project (3GPP) standards.
The performance impact of LoS and NLoS transmissions in SCNs in terms of the coverage probability and the ASE is shown to be significant both quantitatively and qualitatively,
compared with previous work that does not differentiate LoS and NLoS transmissions.
Particularly, our analysis demonstrates that when the density of small cells is larger than a threshold,
the network coverage probability will decrease as small cells become denser,
which in turn makes the ASE suffer from a slow growth or even a notable \emph{decrease}.
For practical regime of small cell density,
the performance results derived from our analysis are distinctively different from previous results,
and  shed new insights on the design and deployment of future dense/ultra-dense SCNs.
\footnote{1536-1276 © 2015 IEEE. Personal use is permitted, but republication/redistribution requires IEEE permission. Please find the final version in IEEE from the link: http://ieeexplore.ieee.org/document/7416981/. Digital Object Identifier: 10.1109/GLOCOM.2015.7416981}
\end{abstract}


\section{Introduction\label{sec:Introduction}}

Dense small cell networks (SCNs)
are considered as the most promising approach to rapidly increase network capacity and meet the ever-increasing capacity demands
in the 5th generation (5G) systems~\cite{Tutor_smallcell}.
However, up to now,  most theoretical studies on SCNs only consider simple path loss models
that do not differentiate Line-of-Sight (LoS) and Non-Line-of-Sight (NLoS) transmissions~{[}2-5{]}.
The major conclusion~{[}2-5{]} is that neither the number of small cells nor the number of cell tiers changes the coverage probability in interference-limited fully-loaded cellular networks.
Such conclusion implies that the area spectral efficiency (ASE) will monotonically grow as small cells go dense.
An intriguing question is: Does this optimistic conclusion still hold when practical LoS and NLoS transmissions are considered in SCNs?

It is well-known that LoS transmission often occurs when the distance between a transmitter and a receiver is small,
while NLoS is more common in long-distance transmissions as well as in office environments and in central business districts.
For a given network environment, when the distance between transmitter and receiver decreases,
the probability that an LoS path exists between them increases,
causing a \emph{transition} from NLoS transmission to LoS transmission.

To the best of authors' knowledge, up to now, performance analysis considering
both LoS and NLoS transmissions
are~\cite{related_work_Jeff} and~\cite{Related_work_Health}.
In~\cite{related_work_Jeff}, the authors assumed a multi-slope piece-wise path loss function.
Such multi-slope piece-wise path loss function does not fit well with the NLoS and LoS model defined by the 3rd Generation Partnership Project (3GPP) standards,
in which the path loss function is not a one-to-one mapping to the distance~\cite{TR36.828}.
In~\cite{Related_work_Health}, the authors treated the event of LoS or NLoS transmission as a probabilistic event for a millimetre wave communication scenario.
To simplify the analysis, the LoS probability function
was approximated by a moment-matched equivalent step function.
The single-piece path loss model and the proposed step function for modeling the transition from NLoS to LoS transmissions
are also not compatible with the model recommended by the 3GPP~[8, 9].

In this paper, we use a general path loss model that features piece-wise path loss functions with probabilistic LoS and NLoS transmissions.
Note that the proposed model is very general and includes almost all existing models used to capture LoS and NLoS transmissions~{[}6-9{]} as its special cases.
The main contributions of the paper are as follows:
\begin{itemize}
\item
Analytical results are obtained for the coverage probability and the ASE
using a general path loss model incorporating both LoS and NLoS transmissions.
\item
Using the above results,
closed-form expressions are further obtained for the coverage probability and the ASE for a special case
based on the 3GPP standards.
\item
Our theoretical analysis reveals an important finding,
i.e., the ASE will initially increase with the increase of the small cell density,
but when the density of small cells becomes sufficiently large,
the network coverage probability will decrease as small cells become denser.
This in turn makes the ASE suffer from a slow growth or even a notable \emph{decrease}.
Thereafter, when the small cell density is very large,
the ASE will then grow almost linearly with the network densification.
These results are not only quantitatively but also qualitatively different from previous study results~{[}2-7{]}.
Thus, our results shed new insights on the
design and deployment of future dense/ultra-dense SCNs in realistic environments.
\end{itemize}

The remainder of this paper is structured as follows.
Section~\ref{sec:System-Model} describes the system model.
Section~\ref{sec:General-Results} presents our main analytical results
on the coverage probability and the ASE,
followed by their application in a 3GPP case in Section~\ref{sec:A-3GPP-Special-Case}.
The derived results are validated using simulations in Section~\ref{sec:Simulation-and-Discussion}.
Finally, the conclusions are drawn in Section~\ref{sec:Conclusion}.

\section{System Model\label{sec:System-Model}}

We consider a DL cellular network in which BSs are deployed in a plane according to a homogeneous Poisson point process (HPPP) $\Phi$ of intensity $\lambda$ $\textrm{BSs/km}^{2}$.
UEs are Poisson distributed in the considered network with an intensity of $\lambda^{\textrm{UE}}$ $\textrm{BSs/km}^{2}$.
$\lambda^{\textrm{UE}}$ is assumed to be sufficiently larger than $\lambda$ so that each BS has at least one associated UE in its coverage area.
The distance between an arbitrary BS and an arbitrary UE is denoted by $r$ in $\textrm{km}$.
Considering practical LoS and NLoS transmissions,
we propose to model the path loss associated with distance $r$ as~(\ref{eq:general_PL_model_our_work}),
shown on the top of the next page.
\begin{algorithm*}[hbtp]
\small
\noindent
\begin{eqnarray}
\small
\zeta\left(r\right) & = & \begin{cases}
\zeta_{1}\left(r\right)=\begin{cases}
\begin{array}{l}
\zeta_{1}^{\textrm{L}}\left(r\right),\\
\zeta_{1}^{\textrm{NL}}\left(r\right),
\end{array} & \begin{array}{l}
\textrm{with probability }\textrm{Pr}_{1}^{\textrm{L}}\left(r\right)\\
\textrm{with probability }\left(1-\textrm{Pr}_{1}^{\textrm{L}}\left(r\right)\right)
\end{array}\end{cases}, & \textrm{when }0\leq r\leq d_{1}\\
\zeta_{2}\left(r\right)=\begin{cases}
\begin{array}{l}
\zeta_{2}^{\textrm{L}}\left(r\right),\\
\zeta_{2}^{\textrm{NL}}\left(r\right),
\end{array} & \begin{array}{l}
\textrm{with probability }\textrm{Pr}_{2}^{\textrm{L}}\left(r\right)\\
\textrm{with probability }\left(1-\textrm{Pr}_{2}^{\textrm{L}}\left(r\right)\right)
\end{array}\end{cases}, & \textrm{when }d_{1}<r\leq d_{2}\\
\vdots & \vdots\\
\zeta_{N}\left(r\right)=\begin{cases}
\begin{array}{l}
\zeta_{N}^{\textrm{L}}\left(r\right),\\
\zeta_{N}^{\textrm{NL}}\left(r\right),
\end{array} & \begin{array}{l}
\textrm{with probability }\textrm{Pr}_{N}^{\textrm{L}}\left(r\right)\\
\textrm{with probability }\left(1-\textrm{Pr}_{N}^{\textrm{L}}\left(r\right)\right)
\end{array}\end{cases}, & \textrm{when }r>d_{N-1}
\end{cases}.\label{eq:general_PL_model_our_work}
\normalsize
\end{eqnarray}
\vspace{-0.2cm}
\end{algorithm*}

In~(\ref{eq:general_PL_model_our_work}),
the path loss function $\zeta\left(r\right)$ is segmented into $N$ pieces with the $N$-th piece $\zeta_{n}\left(r\right)$, $n\in\left\{ 1,2,\ldots,N\right\}$.
For each $\zeta_{n}\left(r\right)$,
$\zeta_{n}^{\textrm{L}}\left(r\right)$ is the $n$-th piece of the path loss function for LoS transmission,
$\zeta_{n}^{\textrm{NL}}\left(r\right)$ is the $n$-th piece of the path loss function for NLoS transmission
and $\textrm{Pr}_{n}^{\textrm{L}}\left(r\right)$ is the $n$-th piece of the LoS probability function.
In more detail,
\begin{itemize}

\item

$\zeta_{n}\left(r\right)$ is modeled as
\vspace{-0.1cm}
\begin{equation}
	\zeta_{n}\left(r\right)=\begin{cases}
	\begin{array}{l}
		\zeta_{n}^{\textrm{L}}\left(r\right)=A_{n}^{{\rm {L}}}r^{-\alpha_{n}^{{\rm {L}}}},\\
		\zeta_{n}^{\textrm{NL}}\left(r\right)=A_{n}^{{\rm {NL}}}r^{-\alpha_{n}^{{\rm {NL}}}},
	\end{array} & \begin{array}{l}
	\textrm{for LoS}\\
	\textrm{for NLoS}
\end{array},\end{cases}\label{eq:PL_BS2UE}
\end{equation}
with $A_{n}^{{\rm {L}}}$ and $A_{n}^{{\rm {NL}}},n\in\left\{ 1,2,\ldots,N\right\} $ being the path losses at a reference distance $r=1$
and $\alpha_{n}^{{\rm {L}}}$ and $\alpha_{n}^{{\rm {NL}}},n\in\left\{ 1,2,\ldots,N\right\} $ being the path loss exponents
for the LoS and the NLoS cases in $\zeta_{n}\left(r\right)$, respectively.
In practice, $A_{n}^{{\rm {L}}}$, $A_{n}^{{\rm {NL}}}$, $\alpha_{n}^{{\rm {L}}}$ and $\alpha_{n}^{{\rm {NL}}}$ are constants obtained from field tests~[8, 9].

\item

$\textrm{Pr}_{n}^{\textrm{L}}\left(r\right)$ is the $n$-th piece probability function that a transmitter and a receiver separated by a distance $r\in[d_{n-1}, d_{n})$ has an LoS path,
which is usually a monotonically decreasing function of $r$.
For convenience, $\left\{ \textrm{Pr}_{n}^{\textrm{L}}\left(r\right)\right\} $ is further stacked into a piece-wise LoS probability function as
\vspace{-0.2cm}
\begin{eqnarray}
	\textrm{Pr}^{\textrm{L}}\left(r\right) & = & \begin{cases} \small
	\textrm{Pr}_{1}^{\textrm{L}}\left(r\right), &  \textrm{when }0\leq r\leq d_{1}\\ \small
	\textrm{Pr}_{2}^{\textrm{L}}\left(r\right), &  \textrm{when }d_{1}<r\leq d_{2}\\ \small
	\vdots & \vdots\\ \small
	\textrm{Pr}_{N}^{\textrm{L}}\left(r\right), &  \textrm{when }r>d_{N-1}
	\end{cases}.\label{eq:general_LoS_Pr_our_work}
\end{eqnarray}

\end{itemize}

Our model is consistent with the ones adopted in the 3GPP~[8, 9].
Note that the considered path loss model in~(\ref{eq:general_PL_model_our_work}) will degenerate to that adopted in~\cite{related_work_Jeff} and~\cite{Related_work_Health}
when $\textrm{Pr}_{n}^{\textrm{L}}\left(r\right)=0,\forall n\in\left\{ 1,2,\ldots,N\right\} $ and $N=1$, respectively.

As a common practice in the field~[2-6],
each UE is assumed to be associated with the nearest BS to the UE,
and the multi-path fading between an arbitrary BS and an arbitrary UE is modeled as independently identical distributed (i.i.d.) Rayleigh fading,
i.e., the channel gain is denoted by $h$ and is modeled as an i.i.d. exponential random variable (RV).
The transmit power of each BS and the additive white Gaussian noise (AWGN) power at each UE are denoted by $P$ and $N_{0}$, respectively.

\section{Analysis Based on General Path Loss Model \label{sec:General-Results}}

Using the properties of the HPPP,
we study the performance of SCNs by considering the performance of a typical UE located at the origin $o$.
We first investigate the coverage probability and thereafter the ASE.

The coverage probability is defined as the probability that the signal to interference plus noise ratio (SINR) of the typical UE, denoted by $\mathrm{SINR}$, is above a threshold $\gamma$:
\vspace{-0.1cm}
\begin{eqnarray}
	p^{\textrm{cov}}\left(\lambda,\gamma\right) & = & \textrm{Pr}\left[\mathrm{SINR}>\gamma\right],\label{eq:Coverage_Prob_def}
\end{eqnarray}

\vspace{-0.1cm}
\noindent where the SINR is computed by
\vspace{-0.2cm}
\begin{eqnarray}
	\mathrm{SINR} & = & \frac{P\zeta\left(r\right)h}{I_{r}+N_{0}},\label{eq:SINR}
\end{eqnarray}

\vspace{-0.1cm}
\noindent where $I_{r}$ is the cumulative interference given by
\vspace{-0.1cm}
\begin{eqnarray}
	I_{r} & = & \sum_{i\in\Phi\setminus{b_{o}}}P\beta_{i}g_{i},\label{eq:cumulative_interference}
\end{eqnarray}

\vspace{-0.1cm}
\noindent where $b_{o}$ is  the BS associated with the typical UE and located at distance $r$ from the typical UE,
and $\beta_{i}$ and $g_{i}$ are the path loss and the multi-path fading channel gain associated with the $i$-th interfering BS, respectively.

According to~\cite{related_work_Jeff} and~\cite{Related_work_Health},
the area spectral efficiency (ASE) in $\textrm{bps/Hz/km}^{2}$ for a given $\lambda$ can be computed by
\vspace{-0.1cm}
\begin{eqnarray}
	A^{\textrm{ASE}}\left(\lambda,\gamma_{0}\right) & = & \lambda\int_{\gamma_{0}}^{\infty}\log_{2}\left(1+x\right)f_{X}\left(\lambda,x\right)dx,\label{eq:ASE_def}
\end{eqnarray}

\vspace{-0.1cm}
\noindent where $\gamma_{0}$ is the minimum working SINR for the considered SCN,
and $f_{X}\left(\lambda,x\right)$ is the probability density function (PDF) of SINR observed at the typical UE at a particular value of $\lambda$.

Since $p^{\textrm{cov}}\left(\lambda,\gamma\right)$ can be defined as the complementary cumulative distribution function (CCDF) of SINR,
$f_{X}\left(\lambda,x\right)$ can be expressed as
\vspace{-0.2cm}
\begin{eqnarray}
	f_{X}\left(\lambda,x\right) & = & \frac{\partial\left(1-p^{\textrm{cov}}\left(\lambda,x\right)\right)}{\partial x}.\label{eq:cond_SINR_PDF}
\end{eqnarray}

\begin{algorithm*}[hbtp]
\small
\begin{thm}
\label{thm:p_cov_UAS2}Considering the path loss model of (\ref{eq:general_PL_model_our_work}), $p^{\textrm{cov}}\left(\lambda,\gamma\right)$ can be
derived as

\noindent
\begin{eqnarray}
	\small
	p^{\textrm{cov}}\left(\lambda,\gamma\right) & = & \sum_{n=1}^{N}\left(T_{n}^{\textrm{L}}+T_{n}^{\textrm{NL}}\right),\label{eq:Theorem_2}
\end{eqnarray}
where $T_{n}^{\textrm{L}}=\int_{d_{n-1}}^{d_{n}}\textrm{Pr}\left[\frac{P\zeta_{n}^{\textrm{L}}\left(r\right)h}{I_{r}+N_{0}}>\gamma\right]f_{R,n}^{\textrm{L}}\left(r\right)dr$,
$T_{n}^{\textrm{NL}}=\int_{d_{n-1}}^{d_{n}}\textrm{Pr}\left[\frac{P\zeta_{n}^{\textrm{NL}}\left(r\right)h}{I_{r}+N_{0}}>\gamma\right]f_{R,n}^{\textrm{NL}}\left(r\right)dr$,
and $d_{0}$ and $d_{N}$ are respectively defined as $0$ and $\infty$.
Moreover, $f_{R,n}^{\textrm{L}}\left(r\right)$ and $f_{R,n}^{\textrm{NL}}\left(r\right)$
are represented as
\noindent
\begin{eqnarray}
	\small
	f_{R,n}^{\textrm{L}}\left(r\right) & = & \textrm{Pr}_{n}^{\textrm{L}}\left(r\right)\times\exp\left(-\pi r^{2}\lambda\right)\times2\pi r\lambda,\quad\left(d_{n-1}<r\leq d_{n}\right),\label{eq:geom_dis_PDF_UAS2_LoS_thm}
\end{eqnarray}
\vspace{-0.2cm}
\noindent and
\noindent
\begin{eqnarray}
	f_{R,n}^{\textrm{NL}}\left(r\right) & = & \left(1-\textrm{Pr}_{n}^{\textrm{L}}\left(r\right)\right)\times\exp\left(-\pi r^{2}\lambda\right)\times2\pi r\lambda,\quad\left(d_{n-1}<r\leq d_{n}\right).		\label{eq:geom_dis_PDF_UAS2_NLoS_thm}
\end{eqnarray}

\noindent Furthermore, $\textrm{Pr}\left[\frac{P\zeta_{n}^{\textrm{L}}\left(r\right)h}{I_{r}+N_{0}}>\gamma\right]$
and $\textrm{Pr}\left[\frac{P\zeta_{n}^{\textrm{NL}}\left(r\right)h}{I_{r}+N_{0}}>\gamma\right]$
are respectively computed by

\noindent
\begin{eqnarray}
\textrm{Pr}\left[\frac{P\zeta_{n}^{\textrm{L}}\left(r\right)h}{I_{r}+N_{0}}>\gamma\right] & = & \exp\left(-\frac{\gamma N_{0}}{P\zeta_{n}^{\textrm{L}}\left(r\right)}\right)\mathscr{L}_{I_{r}}\left(\frac{\gamma}{P\zeta_{n}^{\textrm{L}}\left(r\right)}\right),\label{eq:Pr_SINR_req_UAS2_LoS_thm}
\end{eqnarray}
\vspace{-0.2cm}
\noindent and
\noindent
\begin{eqnarray}
\textrm{Pr}\left[\frac{P\zeta_{n}^{\textrm{NL}}\left(r\right)h}{I_{r}+N_{0}}>\gamma\right] & = & \exp\left(-\frac{\gamma N_{0}}{P\zeta_{n}^{\textrm{NL}}\left(r\right)}\right)\mathscr{L}_{I_{r}}\left(\frac{\gamma}{P\zeta_{n}^{\textrm{NL}}\left(r\right)}\right),\label{eq:Pr_SINR_req_UAS2_NLoS_thm}
\end{eqnarray}
\vspace{-0.2cm}
\noindent where $\mathscr{L}_{I_{r}}\left(s\right)$ is the Laplace
transform of RV $I_{r}$ evaluated at $s$.
\end{thm}

\begin{IEEEproof}
See Appendix~A.
\end{IEEEproof}

\end{algorithm*}
\vspace{-0.1cm}
Given the definition of the coverage probability and the ASE presented in~(\ref{eq:Coverage_Prob_def}) and~(\ref{eq:ASE_def}) respectively,
and using the path loss model of (\ref{eq:general_PL_model_our_work}),
we present our main result on $p^{\textrm{cov}}\left(\lambda,\gamma\right)$
in Theorem~\ref{thm:p_cov_UAS2} shown on the next page.
Plugging $p^{\textrm{cov}}\left(\lambda,\gamma\right)$ from (\ref{eq:Theorem_2}) of Theorem~\ref{thm:p_cov_UAS2} into (\ref{eq:cond_SINR_PDF}),
we can get the ASE using (\ref{eq:ASE_def}).

As can be seen from Theorem~\ref{thm:p_cov_UAS2},
the coverage probability, $p^{\textrm{cov}}\left(\lambda,\gamma\right)$, is a function of
the piece-wise path loss function $\left\{ \zeta_{n}\left(r\right)\right\} $
and the piece-wise LoS probability function $\left\{ \textrm{Pr}_{n}^{\textrm{L}}\left(r\right)\right\} $.
We will investigate their impacts in the sequel.

\section{Study of a 3GPP Special Case \label{sec:A-3GPP-Special-Case}}

As a special case of Theorem~\ref{thm:p_cov_UAS2},
we consider the path loss function, $\zeta\left(r\right)$,
\vspace{-0.1cm}
\begin{equation}
	\zeta\left(r\right)=\begin{cases}
	\begin{array}{l}
		\small
		\hspace{-0.2cm}A^{{\rm {L}}}r^{-\alpha^{{\rm {L}}}},\\
		\hspace{-0.2cm}A^{{\rm {NL}}}r^{-\alpha^{{\rm {NL}}}},
	\end{array} & \begin{array}{l}
	\hspace{-0.5cm}\textrm{w/ probability }\textrm{Pr}^{\textrm{L}}\left(r\right)\\
	\hspace{-0.5cm}\textrm{w/ probability }\left(1-\textrm{Pr}^{\textrm{L}}\left(r\right)\right)
	\end{array}\end{cases},\label{eq:PL_BS2UE_2slopes}
\end{equation}
together with the linear LoS probability function, which is
\vspace{-0.1cm}
\begin{eqnarray}
	\textrm{Pr}^{\textrm{L}}\left(r\right) & = & \begin{cases}
	\begin{array}{l}
		\small
		1-\frac{r}{d_{1}},\\
		0,
		\end{array} & \begin{array}{l}
		0<r\leq d_{1}\\
		r>d_{1}
	\end{array}\end{cases},\label{eq:LoS_Prob_func_linear}
\end{eqnarray}
both respectively recommended in the 3GPP~[8, 9].

Considering the general path loss model presented in~(\ref{eq:general_PL_model_our_work}),
the path loss model presented in~(\ref{eq:PL_BS2UE_2slopes}) and
(\ref{eq:LoS_Prob_func_linear}) can be deemed as a special case of~(\ref{eq:general_PL_model_our_work})
with the following substitution: $N=2$, $\zeta_{1}^{\textrm{L}}\left(r\right)=\zeta_{2}^{\textrm{L}}\left(r\right)=A^{{\rm {L}}}r^{-\alpha^{{\rm {L}}}}$,
$\zeta_{1}^{\textrm{NL}}\left(r\right)=\zeta_{2}^{\textrm{NL}}\left(r\right)=A^{{\rm {NL}}}r^{-\alpha^{{\rm {NL}}}}$,
$\textrm{Pr}_{1}^{\textrm{L}}\left(r\right)=1-\frac{r}{d_{1}}$, and
$\textrm{Pr}_{2}^{\textrm{L}}\left(r\right)=0$. For clarity, this
3GPP special case is referred to as 3GPP Case~1 in the sequel.

According to Theorem~\ref{thm:p_cov_UAS2}, $p^{\textrm{cov}}\left(\lambda,\gamma\right)$ can be obtained as
\vspace{-0.1cm}
\begin{eqnarray}
	p^{\textrm{cov}}\left(\lambda,\gamma\right) & = & \sum_{n=1}^{2}\left(T_{n}^{\textrm{L}}+T_{n}^{\textrm{NL}}\right).
	\label{eq:p_cov_special_case_UAS1}
\end{eqnarray}

In the following subsections, we investigate $T_{1}^{\textrm{L}}$,
$T_{1}^{\textrm{NL}}$, $T_{2}^{\textrm{L}}$, and $T_{2}^{\textrm{NL}}$,
respectively.

\subsection{The Computation of $T_{1}^{\textrm{L}}$ for 3GPP Case~1}

From Theorem~\ref{thm:p_cov_UAS2},
$T_{1}^{\textrm{L}}$ can be derived as
\vspace{-0.1cm}
\begin{eqnarray}
 	T_{1}^{\textrm{L}}=\int_{0}^{d_{1}}\exp\left(-\frac{\gamma r^{\alpha^{{\rm {L}}}}N_{0}}{PA^{{\rm {L}}}}\right)\mathscr{L}_{I_{r}}\left(\frac{\gamma r^{\alpha^{{\rm {L}}}}}{PA^{{\rm {L}}}}\right)f_{R,1}^{\textrm{L}}\left(r\right)dr,		\label{eq:T_1_UAS2_LoS_final}
\end{eqnarray}
where according to Theorem~\ref{thm:p_cov_UAS2} and (\ref{eq:LoS_Prob_func_linear}),
$f_{R,1}^{\textrm{L}}\left(r\right)$ becomes  $f_{R,1}^{\textrm{L}}\left(r\right)=$
\vspace{-0.1cm}
\begin{eqnarray}
	\left(1-\frac{r}{d_{1}}\right)\times\exp\left(-\pi r^{2}\lambda\right)\times2\pi r\lambda,\quad\left(0<r\leq d_{1}\right).\label{eq:spec_geom_dis_PDF_UAS2_LoS_seg1}
\end{eqnarray}

\vspace{-0.3cm}

Furthermore, to compute $\mathscr{L}_{I_{r}}\left(\frac{\gamma r^{\alpha^{{\rm {L}}}}}{PA^{{\rm {L}}}}\right)$
in the range of $0<r\leq d_{1}$, we propose Lemma~\ref{lem:laplace_term_UAS2_LoS_seg1}.
\vspace{-0.2cm}
\begin{lem}
	
	\label{lem:laplace_term_UAS2_LoS_seg1}$\mathscr{L}_{I_{r}}\left(\frac{\gamma r^{\alpha^{{\rm {L}}}}}{PA^{{\rm {L}}}}\right)$
	in the range of $0<r\leq d_{1}$ can be calculated by

	$\mathscr{L}_{I_{r}}\left(\frac{\gamma r^{\alpha^{{\rm {L}}}}}{PA^{{\rm {L}}}}\right)=$
	\vspace{-0.1cm}
	\begin{eqnarray}
		&& \exp\left(-2\pi\lambda\left(\rho_{1}\left(\alpha^{{\rm {L}}},1,\left(\gamma r^{\alpha^{{\rm {L}}}}\right)^{-1},d_{1}\right)\right.\right.\nonumber\\
		&& \left.\left.\qquad\qquad-\rho_{1}\left(\alpha^{{\rm {L}}},1,\left(\gamma r^{\alpha^{{\rm {L}}}}\right)^{-1},r\right)\right)\right)\nonumber \\
		&& \times\exp\left(\frac{2\pi\lambda}{d_{0}}\left(\rho_{1}\left(\alpha^{{\rm {L}}},2,\left(\gamma r^{\alpha^{{\rm {L}}}}\right)^{-1},d_{1}\right)\right.\right.\nonumber\\
		&& \left.\left.\qquad\qquad-\rho_{1}\left(\alpha^{{\rm {L}}},2,\left(\gamma r^{\alpha^{{\rm {L}}}}\right)^{-1},r\right)\right)\right)\nonumber\\
		&& \times\exp\left(-\frac{2\pi\lambda}{d_{0}}\left(\rho_{1}\left(\alpha^{{\rm {NL}}},2,\left(\frac{\gamma A^{{\rm {NL}}}}{A^{{\rm {L}}}}r^{\alpha^{{\rm {L}}}}\right)^{-1},d_{1}\right)\right.\right.\nonumber\\
		&& \left.\left.\qquad\qquad-\rho_{1}\left(\alpha^{{\rm {NL}}},2,\left(\frac{\gamma A^{{\rm {NL}}}}{A^{{\rm {L}}}}r^{\alpha^{{\rm {L}}}}\right)^{-1},r\right)\right)\right)\nonumber\\
		&& \times\exp\left(-2\pi\lambda\rho_{2}\left(\alpha^{{\rm {NL}}},1,\left(\frac{\gamma A^{{\rm {NL}}}}{A^{{\rm {L}}}}r^{\alpha^{{\rm {L}}}}\right)^{-1},d_{1}\right)\right),\nonumber \\
		&&\qquad\qquad\qquad\qquad\qquad\qquad\qquad\left(0<r\leq d_{1}\right),\label{eq:Lemma_6}
	\end{eqnarray}

	\vspace{-0.3cm}
	\noindent where

	\noindent $\rho_{1}\left(\alpha,\beta,t,d\right) = $
	\begin{eqnarray}
\left[\frac{d^{\left(\beta+1\right)}}{\beta+1}\right]{}_{2}F_{1}\left[1,\frac{\beta+1}{\alpha};1+\frac{\beta+1}{\alpha};-td^{\alpha}\right],\label{eq:rou1_func}
	\end{eqnarray}

	\vspace{-0.3cm}
	\noindent and

	\noindent $\rho_{2}\left(\alpha,\beta,t,d\right) = $
	\begin{eqnarray}
		\left[\frac{d^{-\left(\alpha-\beta-1\right)}}{t\left(\alpha-\beta-1\right)}\right]{}_{2}F_{1}\left[1,1-\frac{\beta+1}{\alpha};2-\frac{\beta+1}{\alpha};-\frac{1}{td^{\alpha}}\right],\label{eq:rou2_func}
	\end{eqnarray}
\noindent where $_{2}F_{1}\left[\cdot,\cdot;\cdot;\cdot\right]$ is
the hyper-geometric function~\cite{Book_Integrals}.
\end{lem}
\vspace{-0.1cm}
\begin{IEEEproof}
See Appendix~B.
\end{IEEEproof}

\subsection{The Computation of $T_{1}^{\textrm{NL}}$ for 3GPP Case~1}

From Theorem~\ref{thm:p_cov_UAS2},
$T_{1}^{\textrm{NL}}$ can be derived as
\vspace{-0.1cm}
\begin{eqnarray}
	T_{1}^{\textrm{NL}}=\hspace{-0.2cm}\int_{0}^{d_{1}}\hspace{-0.2cm}\exp\left(-\frac{\gamma r^{\alpha^{{\rm {NL}}}}N_{0}}{PA^{{\rm {NL}}}}\right)\mathscr{L}_{I_{r}}\left(\frac{\gamma r^{\alpha^{{\rm {NL}}}}}{PA^{{\rm 	{NL}}}}\right)\hspace{-0.1cm}f_{R,1}^{\textrm{NL}}\left(r\right)dr,\label{eq:T_1_UAS2_NLoS_final}
\end{eqnarray}
where according to Theorem~\ref{thm:p_cov_UAS2} and (\ref{eq:LoS_Prob_func_linear}),
$f_{R,1}^{\textrm{NL}}\left(r\right)$ becomes
\vspace{-0.1cm}
\begin{eqnarray}
	f_{R,1}^{\textrm{NL}}\left(r\right)=\frac{r}{d_{1}}\times\exp\left(-\pi r^{2}\lambda\right)\times2\pi r\lambda,\quad\left(0<r\leq d_{1}\right).\label{eq:spec_geom_dis_PDF_UAS2_NLoS_seg1}
\end{eqnarray}

\vspace{-0.3cm}

Furthermore, to compute $\mathscr{L}_{I_{r}}\left(\frac{\gamma r^{\alpha^{{\rm {NL}}}}}{PA^{{\rm {NL}}}}\right)$
in the range of $0<r\leq d_{1}$, we propose Lemma~\ref{lem:laplace_term_UAS2_NLoS_seg1}.
\vspace{-0.2cm}
\begin{lem}
	
	\noindent \label{lem:laplace_term_UAS2_NLoS_seg1}$\mathscr{L}_{I_{r}}\left(\frac{\gamma r^{\alpha^{{\rm {NL}}}}}{PA^{{\rm {NL}}}}\right)$
	in the range of $0<r\leq d_{1}$ can be calculated by

	$\mathscr{L}_{I_{r}}\left(\frac{\gamma r^{\alpha^{{\rm {NL}}}}}{PA^{{\rm {NL}}}}\right) = $
	\vspace{-0.2cm}
	\begin{eqnarray}
		&& \exp\left(-2\pi\lambda\left(\rho_{1}\left(\alpha^{{\rm {L}}},1,\left(\frac{\gamma A^{{\rm {L}}}}{A^{{\rm {NL}}}}r^{\alpha^{{\rm {NL}}}}\right)^{-1},d_{1}\right)\right.\right.\nonumber\\
		&& \left.\left.\qquad\qquad-\rho_{1}\left(\alpha^{{\rm {L}}},1,\left(\frac{\gamma A^{{\rm {L}}}}{A^{{\rm {NL}}}}r^{\alpha^{{\rm {NL}}}}\right)^{-1},r\right)\right)\right)\nonumber \\
		&& \times\exp\left(\frac{2\pi\lambda}{d_{0}}\left(\rho_{1}\left(\alpha^{{\rm {L}}},2,\left(\frac{\gamma A^{{\rm {L}}}}{A^{{\rm {NL}}}}r^{\alpha^{{\rm {NL}}}}\right)^{-1},d_{1}\right)\right.\right.\nonumber\\
		&& \left.\left.\qquad\qquad-\rho_{1}\left(\alpha^{{\rm {L}}},2,\left(\frac{\gamma A^{{\rm {L}}}}{A^{{\rm {NL}}}}r^{\alpha^{{\rm {NL}}}}\right)^{-1},r\right)\right)\right)\nonumber\\
		&& \times\exp\left(-\frac{2\pi\lambda}{d_{0}}\left(\rho_{1}\left(\alpha^{{\rm {NL}}},2,\left(\gamma r^{\alpha^{{\rm {NL}}}}\right)^{-1},d_{1}\right)\right.\right.\nonumber\\
		&& \left.\left.\qquad\qquad-\rho_{1}\left(\alpha^{{\rm {NL}}},2,\left(\gamma r^{\alpha^{{\rm {NL}}}}\right)^{-1},r\right)\right)\right)\nonumber\\
		&& \times\exp\left(-2\pi\lambda\rho_{2}\left(\alpha^{{\rm {NL}}},1,\left(\gamma r^{\alpha^{{\rm {NL}}}}\right)^{-1},d_{1}\right)\right),\nonumber \\
		&&\qquad\qquad\qquad\qquad\qquad\qquad\qquad\left(0<r\leq d_{1}\right),\label{eq:Lemma_7}
	\end{eqnarray}

	\vspace{-0.2cm}
	\noindent where $\rho_{1}\left(\alpha,\beta,t,d\right)$ and $\rho_{2}\left(\alpha,\beta,t,d\right)$ are defined in (\ref{eq:rou1_func}) and (\ref{eq:rou2_func}), respectively.
\end{lem}
\vspace{-0.1cm}
\begin{IEEEproof}
	See Appendix~C.
\end{IEEEproof}

\subsection{The Computation of $T_{2}^{\textrm{L}}$ for 3GPP Case~1}

From Theorem~\ref{thm:p_cov_UAS2},
$T_{2}^{\textrm{L}}$ can be derived as
\vspace{-0.2cm}
\begin{eqnarray}
	T_{2}^{\textrm{L}} & = & \int_{d_{1}}^{\infty}\exp\left(-\frac{\gamma r^{\alpha^{{\rm {L}}}}N_{0}}{PA^{{\rm {L}}}}\right)\mathscr{L}_{I_{r}}\left(\frac{\gamma r^{\alpha^{{\rm {L}}}}}{PA^{{\rm {L}}}}\right)f_{R,2}^{\textrm{L}}\left(r	\right)dr\nonumber \\
 	& = & 0.\label{eq:T_2_UAS2_LoS_final}
\end{eqnarray}

\vspace{-0.3cm}

Note that the reason why $T_{2}^{\textrm{L}}=0$ in (\ref{eq:T_2_UAS2_LoS_final}) is because according to Theorem~\ref{thm:p_cov_UAS2} and (\ref{eq:LoS_Prob_func_linear}),
we have
\vspace{-0.1cm}
\begin{eqnarray}
	f_{R,2}^{\textrm{L}}\left(r\right) & = & 0\times\exp\left(-\pi r^{2}\lambda\right)\times2\pi r\lambda\nonumber \\
	 & = & 0,\quad\left(r>d_{1}\right).\label{eq:spec_geom_dis_PDF_UAS2_LoS_seg2}
\end{eqnarray}

\subsection{The Computation of $T_{2}^{\textrm{NL}}$ for 3GPP Case~1}

From Theorem~\ref{thm:p_cov_UAS2},
$T_{2}^{\textrm{NL}}$ can be derived as
\vspace{-0.1cm}
\begin{eqnarray}
	T_{2}^{\textrm{NL}}=\hspace{-0.2cm}\int_{d_{1}}^{\infty}\hspace{-0.2cm}\exp\left(-\frac{\gamma r^{\alpha^{{\rm {NL}}}}N_{0}}{PA^{{\rm {NL}}}}\right)\mathscr{L}_{I_{r}}\left(\frac{\gamma r^{\alpha^{{\rm {NL}}}}}{PA^{{\rm {NL}}}}\right)\hspace{-0.1cm}f_{R,2}^{\textrm{NL}}\left(r\right)dr,\label{eq:T_2_UAS2_NLoS_final}
\end{eqnarray}
where according to Theorem~\ref{thm:p_cov_UAS2} and (\ref{eq:LoS_Prob_func_linear}),
$f_{R,2}^{\textrm{NL}}\left(r\right)$ becomes
\vspace{-0.1cm}
\begin{eqnarray}
	f_{R,2}^{\textrm{NL}}\left(r\right) & = & 1\times\exp\left(-\pi r^{2}\lambda\right)\times2\pi r\lambda\nonumber \\
 	& = & \exp\left(-\pi\lambda r^{2}\right)\times2\pi r\lambda,\quad\left(r>d_{1}\right).\label{eq:spec_geom_dis_PDF_UAS2_NLoS_seg2}
\end{eqnarray}

\vspace{-0.2cm}

Furthermore, to compute $\mathscr{L}_{I_{r}}\left(\frac{\gamma r^{\alpha^{{\rm {NL}}}}}{PA^{{\rm {NL}}}}\right)$
in the range of $r>d_{1}$, we propose Lemma~\ref{lem:laplace_term_UAS1_NLoS_seg2}.
\vspace{-0.1cm}

\begin{lem}

	
	\label{lem:laplace_term_UAS1_NLoS_seg2}
$\mathscr{L}_{I_{r}}\left(\frac{\gamma r^{\alpha^{{\rm {NL}}}}}{PA^{{\rm {NL}}}}\right)$ in the range of $r>d_{1}$ can be calculated by

	\noindent $\mathscr{L}_{I_{r}}\left(\frac{\gamma r^{\alpha^{{\rm {NL}}}}}{PA^{{\rm {NL}}}}\right) = $
	\vspace{-0.2cm}
	\begin{eqnarray}
		\hspace{0.3cm}\exp\left(\hspace{-0.1cm}-2\pi\lambda\rho_{2}\left(\alpha^{{\rm {NL}}},1,\left(\gamma r^{\alpha^{{\rm {NL}}}}\right)^{-1},r\right)\hspace{-0.1cm}\right),\;\left(r>d_{1}\right),\label{eq:Lemma_5}
	\end{eqnarray}

	\noindent where $\rho_{2}\left(\alpha,\beta,t,d\right)$ is defined in (\ref{eq:rou2_func}).
\end{lem}
\vspace{-0.1cm}
\begin{IEEEproof}
	See Appendix~D.
\end{IEEEproof}

\subsection{The Results of $p^{\textrm{cov}}\left(\lambda,\gamma\right)$ and
$A^{\textrm{ASE}}\left(\lambda,\gamma_{0}\right)$}

To sum up,
$p^{\textrm{cov}}\left(\lambda,\gamma\right)$ for 3GPP
Case~1 can be written as
\vspace{-0.1cm}
\begin{eqnarray}
	p^{\textrm{cov}}\left(\lambda,\gamma\right) & = & T_{1}^{\textrm{L}}+T_{1}^{\textrm{NL}}+T_{2}^{\textrm{NL}},\label{eq:spec_p_cov_UAS2_final}
\end{eqnarray}
where $T_{1}^{\textrm{L}}$, $T_{1}^{\textrm{NL}}$ and $T_{2}^{\textrm{NL}}$ are computed from closed-form expressions using
(\ref{eq:T_1_UAS2_LoS_final}), (\ref{eq:T_1_UAS2_NLoS_final}) and (\ref{eq:T_2_UAS2_NLoS_final}), respectively.

Plugging $p^{\textrm{cov}}\left(\lambda,\gamma\right)$ obtained from (\ref{eq:spec_p_cov_UAS2_final}) into (\ref{eq:cond_SINR_PDF}),
we can get the result of $A^{\textrm{ASE}}\left(\lambda,\gamma_{0}\right)$ from (\ref{eq:ASE_def}) for 3GPP Case~1.

\section{Simulation and Discussion\label{sec:Simulation-and-Discussion}}

\begin{figure*}[htbp]
\begin{minipage}{0.47\linewidth}
\centering
\includegraphics[width=7.5cm]{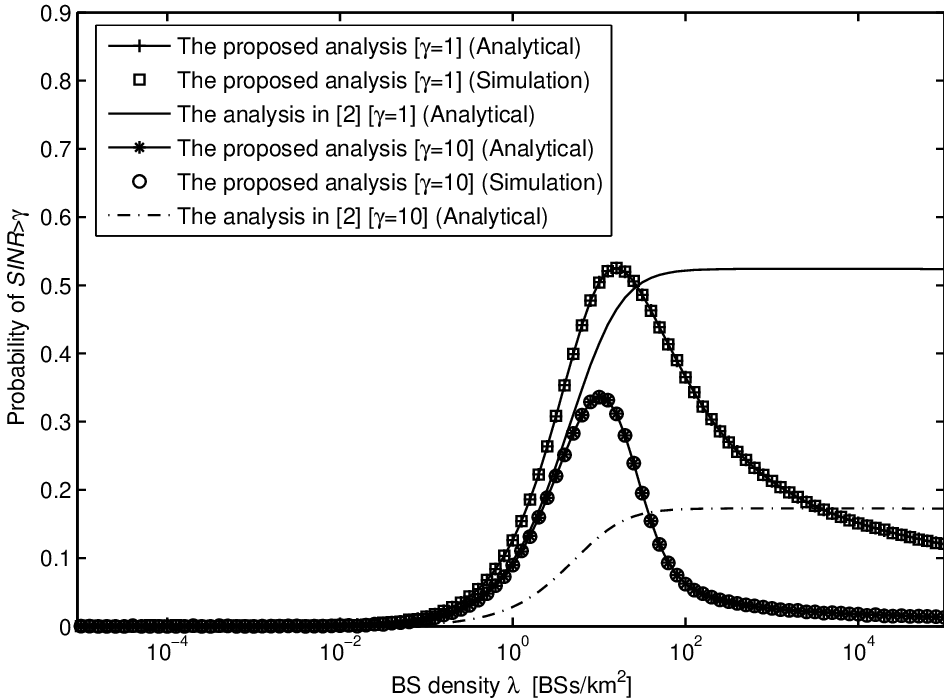}\renewcommand{\figurename}{Fig.}\protect\caption{\label{fig:p_cov_3gpp_linear}$p^{\textrm{cov}}\left(\lambda,\gamma\right)$
vs $\lambda$ for 3GPP Case~1.}
\vspace{-0.2cm}
\end{minipage}
\hspace{0.5cm}
\begin{minipage}{0.47\linewidth}
\centering
\includegraphics[width=7.6cm]{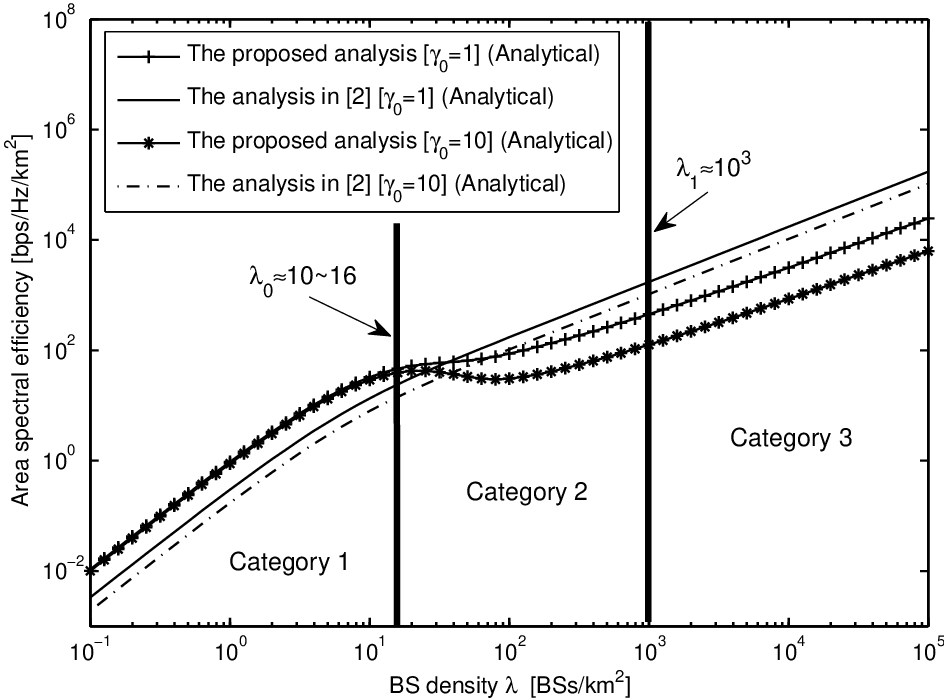}\renewcommand{\figurename}{Fig.}\protect\caption{\label{fig:ASE_3gpp_linear}$A^{\textrm{ASE}}\left(\lambda,\gamma_0\right)$
vs $\lambda$ for 3GPP Case~1.}
\vspace{-0.2cm}
\end{minipage}
\vspace{-0.3cm}
\end{figure*}

In this Section,
we use simulations to further study the performance of SCNs and establish the accuracy of our analysis for 3GPP Case~1 studied in Section~\ref{sec:A-3GPP-Special-Case}.
According to~\cite{TR36.828} and~\cite{SCM_pathloss_model},
we use the following parameters:
$d_{1}=0.3$\,km, $\alpha^{{\rm {L}}}=2.09$, $\alpha^{{\rm {NL}}}=3.75$, $A^{{\rm {L}}}=10^{-4.11}$, $A^{{\rm {NL}}}=10^{-3.29}$, $P=24$\,dBm, $N_{0}=-95$\,dBm.

\subsection{Validation and Discussion of $p^{\textrm{cov}}\left(\lambda,\gamma\right)$\label{sub:Sim-Verification_p_cov}}

The results of $p^{\textrm{cov}}\left(\lambda,\gamma\right)$ with $\gamma=1$ and $\gamma=10$ are plotted in
Fig.~\ref{fig:p_cov_3gpp_linear}.
As can be observed from both figures,
our analytical results perfectly match the simulation results.
Since the results of $A^{\textrm{ASE}}\left(\lambda,\gamma_{0}\right)$ are computed based on $p^{\textrm{cov}}\left(\lambda,\gamma\right)$,
we will only use analytical results for $p^{\textrm{cov}}\left(\lambda,\gamma\right)$ in our discussion hereafter.
For comparison, we have also included analytical results assuming a simplistic path loss model that does not differentiate LoS and NLoS transmissions~\cite{Jeff's work 2011}.
Note that in~\cite{Jeff's work 2011},
only one path loss exponent is defined and denoted by $\alpha$.
Here, $\alpha$ is set to
$\alpha^{{\rm {NL}}}$
to show the analytical results from~\cite{Jeff's work 2011}.

From Fig.~\ref{fig:p_cov_3gpp_linear},
we can observe that
the coverage probability performance given by the stochastic geometry analysis in~\cite{Jeff's work 2011} first increases with the BS density
because more BSs provide better coverage in noise-limited networks.
Then, when $\lambda$ is large enough,
the coverage probability becomes independent of $\lambda$ since the network is pushed into the interference-limited region,
e.g.,
$\lambda>10^{2}\,\textrm{BSs/km}^{2}$ for the analysis in~\cite{Jeff's work 2011}. 
This observation is consistent with the conclusion in~\cite{Jeff's work 2011},
which shows that for a sufficiently large $\lambda$,
the coverage probability becomes almost a constant with the increase of the small cell density.
The intuition behind the observation is that with the simplistic assumption on the path loss model,
the increase in interference power is counterbalanced by the increase in signal power in a interference-limited network,
and thus the coverage probability remains the same as $\lambda$ increases.

\vspace{-0.05cm}
In Fig.~\ref{fig:p_cov_3gpp_linear},
the coverage probability performance of the proposed stochastic geometry analysis
incorporating both LoS and NLoS transmissions exhibits a significant deviation from that of the analysis from~\cite{Jeff's work 2011},
because when the distance $r$ decreases, or equivalently when the small cell density $\lambda$ increases,
LoS transmission occurs with an increasingly higher probability than NLoS transmission.
Specifically, when the SCN is sparse and thus noise-limited,
e.g., $\lambda\leq10\,\textrm{BSs/km}^{2}$,
the coverage probability given by the proposed analysis grows as $\lambda$ increases for the same reason as explained in the above paragraph,
i.e., deploying more small cells is beneficial for removing coverage holes.
Then, when the network is dense enough and all coverage holes are removed,
the coverage probability given by the proposed analysis decreases as $\lambda$ increases,
due to the transition of a large number of interference paths from NLoS to LoS.
It is important to note that the coverage probability performance of the proposed analysis for 3GPP Case~1 peaks at a certain value $\lambda_{0}$.
When $\lambda$ increases above $\lambda_{0}$,
interfering BSs may be very close to the typical UE and hence their signals may reach the UE via strong LoS paths.
Such critical point of $\lambda_{0}$ can be readily obtained by setting the partial derivative of $p^{\textrm{cov}}\left(\lambda,\gamma\right)$ with regard to $\lambda$ to zero,
i.e., $\lambda_{0}=\underset{\lambda}{\arg}\left\{ \frac{\partial p^{\textrm{cov}}\left(\lambda,\gamma\right)}{\partial\lambda}=0\right\} $.
The solution to this equation can be numerically found using a standard bisection searching~\cite{Bisection}.
In Fig.~\ref{fig:p_cov_3gpp_linear},
the numerical results of $\lambda_{0}$ are 15.85\,$\textrm{BSs/km}^{2}$ and 10.21\,$\textrm{BSs/km}^{2}$ for $\gamma=1$ and $\gamma=10$, respectively.

\subsection{Discussion of the Analytical Results of $A^{\textrm{ASE}}\left(\lambda,\gamma_{0}\right)$\label{sub:Investigation-ASE}}

The results of $A^{\textrm{ASE}}\left(\lambda,1\right)$ and $A^{\textrm{ASE}}\left(\lambda,10\right)$ are plotted in Fig.~\ref{fig:ASE_3gpp_linear},
comparing the proposed stochastic geometry analysis with the conventional stochastic geometry analysis in~\cite{Jeff's work 2011}.

As can be seen from Fig.~\ref{fig:ASE_3gpp_linear},
the analysis from~\cite{Jeff's work 2011} indicates that when the SCN is dense enough,
e.g., $\lambda\geq10^{2}\,\textrm{BSs/km}^{2}$,
the ASE performance increases almost linearly with $\lambda$,
which is logically correct from the conclusion that $p^{\textrm{cov}}\left(\lambda,\gamma\right)$ is invariable with respect to $\lambda$ for a given $\gamma$ in dense SCNs~\cite{Jeff's work 2011}.

In contrast, the proposed stochastic geometry analysis
reveals a more complicated trend for the ASE performance.
Specifically, when the SCN is relatively sparse,
e.g., $\lambda\leq\lambda_{0}$,
the ASE quickly increases with $\lambda$ because the network is generally noise-limited,
thus adding more small cells immensely benefits the ASE.
When the SCN is extremely dense,
e.g., $\lambda\geq10^{4}\,\textrm{BSs/km}^{2}$,
the ASE exhibits a nearly linear trajectory with regard to $\lambda$
because both the signal power and the interference power are now LoS dominated and thus statistically stable.
Note that the pace is lower than in~\cite{Jeff's work 2011}.
As for the practical range of $\lambda$, i.e., $\lambda\in\left[\lambda_{0},10^{4}\right]\,\textrm{BSs/km}^{2}$,
the ASE first exhibits a slowing-down in the rate of growth (when $\gamma=1$) or even a notable \emph{decrease} in its absolute value (when $\gamma=10$).
This is attributed to the fast decrease of the coverage probability at around $\lambda\in\left[\lambda_{0},\lambda_{1}\right]\,\textrm{BSs/km}^{2}$
as shown in Fig.~\ref{fig:p_cov_3gpp_linear},
where $\lambda_{1}$ is another threshold larger than $\lambda_{0}$.
When $\lambda\geq\lambda_{1}$,
the ASE will pick up the growth rate as the decrease of the coverage probability becomes very gentle.
In Fig.~\ref{fig:ASE_3gpp_linear},
the value of $\lambda_{1}$ seems to be around $10^{3}\,\textrm{BSs/km}^{2}$.

Our new finding indicates the significant impact of the path loss model incorporating both NLoS and LoS transmissions.
As a confirmation,
noting that in Fig.~\ref{fig:p_cov_3gpp_linear},
we can observe that increasing $\gamma$ from 1 to 10 will greatly accelerate the decrease of the coverage probability at around $\lambda\in\left[10,10^{2}\right]\,\textrm{BSs/km}^{2}$,
which in turn causes the notable \emph{decrease} of the ASE at that range of $\lambda$ when $\gamma=10$ in Fig.~\ref{fig:ASE_3gpp_linear}.



With the defined thresholds $\lambda_{0}$ and $\lambda_{1}$,
SCNs can be roughly classified into 3 categories,
i.e., the sparse SCN ($0<\lambda\leq\lambda_{0}$), the dense SCN ($\lambda_{0}<\lambda\leq\lambda_{1}$) and the very dense SCN ($\lambda>\lambda_{1}$).
The ASEs for both the sparse SCN and the very dense SCN grow almost linearly with the increase of $\lambda$,
while the ASE of the dense SCN shows a slow growth or even a notable \emph{decrease} with the increase of $\lambda$.
From Fig.~\ref{fig:ASE_3gpp_linear},
we can get a new look at the ultra-dense SCN,
which has been identified as one of the key enabling technologies of the 5G networks~\cite{Tutor_smallcell}.
Up to now, there is no consensus in both industry and academia on that at what density a SCN can be categorized as an ultra-dense SCN.
According to our study,
for 3GPP Case~1, we propose that the 5G systems should target the third category of SCNs as ultra-dense SCNs,
i.e., the SCNs with $\lambda>\lambda_{1}$,
because the associated ASE will grow almost linearly as $\lambda$ increases.
Numerically speaking, $\lambda_{1}$ is around $10^{3}\,\textrm{BSs/km}^{2}$
from Fig.~\ref{fig:ASE_3gpp_linear}.
It is particularly important to note that the second category of SCNs ($\lambda_{0}<\lambda\leq\lambda_{1}$) is better avoided in practical SCN deployments due to its cost-inefficiency shown in Fig.~\ref{fig:ASE_3gpp_linear}.

\section{Conclusion\label{sec:Conclusion}}

In this paper, we have shown that a sophisticated path loss model incorporating both LoS and NLoS transmissions has a significant impact on the performance of SCNs,
measured by the two metrics of the coverage probability and the ASE.
Such impact is not only quantitative but also qualitative.
Specifically, our theoretical analysis have concluded that the ASE will initially increase with the increase of the small cell density,
but when the density of small cells is larger than a threshold $\lambda_{0}$,
the network coverage probability will decrease,
which in turn makes the ASE suffer from a slow growth or even a notable \emph{decrease} as the small cell density increases.
Furthermore, the ASE will grow almost linearly as the small cell density increases above another larger threshold $\lambda_{1}$.
According to our study, for 3GPP cases,
we propose that the 5G systems should target the SCNs with $\lambda>\lambda_{1}$ as ultra-dense SCNs.
Numerically speaking, $\lambda_{1}$ appears to be around several $10^{3}\,\textrm{BSs/km}^{2}$.

The intuition behind our conclusion is that when the density of small cells is larger than a threshold,
the interference power will increase faster than the signal power due to the transition of a large number of interference paths from NLoS to LoS,
and thus the small cell density matters!

As our future work, we will incorporate more sophisticated UE association strategies and more practical multi-path fading model
into the analysis of SCNs because the multi-path fading model is also
affected by LoS and NLoS transmissions.

\section*{Appendix~A: Proof of Theorem~\ref{thm:p_cov_UAS2} \label{sec:Appendix-A}}

%

From (\ref{eq:Coverage_Prob_def}) and (\ref{eq:SINR}), we can derive
$p^{\textrm{cov}}\left(\lambda,\gamma\right)$ straightforwardly as
\vspace{-0.3cm}
\noindent
\begin{eqnarray}
p^{\textrm{cov}}\left(\lambda,\gamma\right) & \stackrel{\left(a\right)}{=} & \int_{r>0}\textrm{Pr}\left[\left.\mathrm{SINR}>\gamma\right|r\right]f_{R}\left(r\right)dr\nonumber \\
 & = & \int_{r>0}\textrm{Pr}\left[\left.\frac{P\zeta\left(r\right)h}{I_{r}+N_{0}}>\gamma\right|r\right]f_{R}\left(r\right)dr\nonumber \\
 & \stackrel{\bigtriangleup}{=} & \sum_{n=1}^{N}\left(T_{n}^{\textrm{L}}+T_{n}^{\textrm{NL}}\right),\label{eq:p_cov_general_form}
\end{eqnarray}
\normalsize

\noindent where $T_{n}^{\textrm{L}}$ and $T_{n}^{\textrm{NL}}$ are
piece-wise functions defined as $T_{n}^{\textrm{L}}=\int_{d_{n-1}}^{d_{n}}\textrm{Pr}\left[\frac{P\zeta_{n}^{\textrm{L}}\left(r\right)h}{I_{r}+N_{0}}>\gamma\right]f_{R,n}^{\textrm{L}}\left(r\right)dr$
and $T_{n}^{\textrm{NL}}=\int_{d_{n-1}}^{d_{n}}\textrm{Pr}\left[\frac{P\zeta_{n}^{\textrm{NL}}\left(r\right)h}{I_{r}+N_{0}}>\gamma\right]f_{R,n}^{\textrm{NL}}\left(r\right)dr$,
respectively. Besides, $d_{0}$ and $d_{N}$ are respectively defined
as $0$ and $\infty$. Moreover, $f_{R,n}^{\textrm{L}}\left(r\right)$
and $f_{R,n}^{\textrm{NL}}\left(r\right)$ are the piece-wise PDFs
of the event that the UE is associated with the nearest BS with an LoS path
at distance $r$ and the event that the UE is associated
with the nearest BS with an NLoS path at distance $r$, respectively.

Regarding $f_{R,n}^{\textrm{L}}\left(r\right)$, we define two events
in the following, whose joint event
is equivalent to the event that
the UE is associated with a BS with an LoS path at distance $r$.
\begin{itemize}
\item Event $B$: the nearest BS is located at distance $r$
\item Event $D^{\textrm{L}}$: the BS is one with an LoS path
\end{itemize}
According to~\cite{Jeff's work 2011}, the cumulative density function
(CDF) of Event $B$ with regard to $r$ is given by
\vspace{-0.1cm}
\noindent
\begin{eqnarray}
F_{R,n}^{B}\left(r\right)=1-\exp\left(-\pi r^{2}\lambda\right),\quad\left(d_{n-1}<r\leq d_{n}\right).\label{eq:geom_dis_CDF_UAS2_LoS_eventB}
\end{eqnarray}

\noindent Hence, taking the derivative of $F_{R,n}^{B}\left(r\right)$
with regard to $r$, yields the PDF of Event $B$ as
\vspace{-0.1cm}
\noindent
\begin{eqnarray}
f_{R,n}^{B}\left(r\right)=\exp\left(-\pi r^{2}\lambda\right)\times2\pi r\lambda,\quad\left(d_{n-1}<r\leq d_{n}\right).\label{eq:geom_dis_PDF_UAS2_LoS_eventB}
\end{eqnarray}

\noindent The PDF $f_{R,n}^{B}\left(r\right)$ should be further thinned
by the probability of Event $D^{\textrm{L}}$ on condition of $r$,
which is $\textrm{Pr}_{n}^{\textrm{L}}\left(r\right)$, so that we
can get the PDF of the joint event of $B$ and $C^{\textrm{NL}}$
as
\vspace{-0.1cm}
\noindent
\begin{eqnarray}
f_{R,n}^{\textrm{L}}\left(r\right)=\textrm{Pr}_{n}^{\textrm{L}}\left(r\right)\times f_{R,n}^{B}\left(r\right).\label{eq:geom_dis_PDF_UAS2_LoS}
\end{eqnarray}

Regarding $f_{R,n}^{\textrm{NL}}\left(r\right)$, we also define two
events
in the following, whose joint event
is equivalent to the event
that the UE is associated with a BS with an NLoS path at distance
$r$.
\begin{itemize}
\item Event $B$: the nearest BS is located at distance $r$
\item Event $D^{\textrm{NL}}$: the BS is one with an NLoS path
\end{itemize}
\noindent Similar to (\ref{eq:geom_dis_PDF_UAS2_LoS}), the PDF $f_{R,n}^{B}\left(r\right)$
should be further thinned by the probability of Event $D^{\textrm{L}}$
on condition of $r$, which is $\left(1-\textrm{Pr}_{n}^{\textrm{L}}\left(r\right)\right)$,
so that we can get the PDF of the joint event of $B$ and $D^{\textrm{L}}$
as

\noindent
$f_{R,n}^{\textrm{NL}}\left(r\right) = $
\vspace{-0.1cm}
\begin{eqnarray}
\hspace{0.2cm}\left(1-\textrm{Pr}_{n}^{\textrm{L}}\left(r\right)\right)\hspace{-0.1cm}\times\hspace{-0.1cm}\exp\left(-\pi r^{2}\lambda\right)\hspace{-0.1cm}\times\hspace{-0.1cm}2\pi r\lambda,\left(d_{n-1}\hspace{-0.1cm}<r\hspace{-0.1cm}\leq d_{n}\right).\label{eq:geom_dis_PDF_UAS2_NLoS}
\end{eqnarray}

As for the calculation of $\textrm{Pr}\left[\frac{P\zeta_{n}^{\textrm{L}}\left(r\right)h}{I_{r}+N_{0}}>\gamma\right]$
in~(\ref{eq:p_cov_general_form}), we have

\noindent
\begin{eqnarray}
\textrm{Pr}\left[\frac{P\zeta_{n}^{\textrm{L}}\left(r\right)h}{I_{r}+N_{0}}>\gamma\right] & \hspace{-0.3cm}= \hspace{-0.3cm}& \mathbb{E}_{\left[I_{r}\right]}\left\{ \textrm{Pr}\left[h>\frac{\gamma\left(I_{r}+N_{0}\right)}{P\zeta_{n}^{\textrm{L}}\left(r\right)}\right]\right\} \nonumber \\
 & \hspace{-0.3cm}= \hspace{-0.3cm}& \mathbb{E}_{\left[I_{r}\right]}\left\{ \bar{F}_{H}\left(\frac{\gamma\left(I_{r}+N_{0}\right)}{P\zeta_{n}^{\textrm{L}}\left(r\right)}\right)\right\} ,\label{eq:Pr_SINR_req_UAS1_LoS}
\end{eqnarray}
\normalsize

\noindent where $\mathbb{E}_{\left[X\right]}\left\{ \cdot\right\} $
denotes the expectation operation by taking the expectation over the
variable $X$ and $\bar{F}_{H}\left(h\right)$ denotes the complementary
cumulative density function (CCDF) of RV $h$. Since we assume $h$
to be an exponential RV, we have $\bar{F}_{H}\left(h\right)=\exp\left(-h\right)$
and thus~(\ref{eq:Pr_SINR_req_UAS1_LoS})
can be further derived as

\noindent
\begin{eqnarray}
\textrm{Pr}\left[\frac{P\zeta_{n}^{\textrm{L}}\left(r\right)h}{I_{r}+N_{0}}>\gamma\right]&\hspace{-0.4cm}=\hspace{-0.3cm}&\mathbb{E}_{\left[I_{r}\right]}\left\{ \exp\left(-\frac{\gamma\left(I_{r}+N_{0}\right)}{P\zeta_{n}^{\textrm{L}}\left(r\right)}\right)\right\} \hspace{1.2cm}\nonumber \\
&\hspace{-0.4cm}=\hspace{-0.3cm}&\exp\hspace{-0.1cm}\left(-\frac{\gamma N_{0}}{P\zeta_{n}^{\textrm{L}}\left(r\right)}\right)\hspace{-0.1cm}\mathscr{L}_{I_{r}}\hspace{-0.1cm}\left(\frac{\gamma}{P\zeta_{n}^{\textrm{L}}\left(r\right)}\right),\label{eq:Pr_SINR_req_wLT_UAS1_LoS}
\end{eqnarray}

\noindent where $\mathscr{L}_{I_{r}}\left(s\right)$ is the Laplace
transform of $I_{r}$ evaluated at $s$.

As for the calculation of $\textrm{Pr}\left[\frac{P\zeta_{n}^{\textrm{NL}}\left(r\right)h}{I_{r}+N_{0}}>\gamma\right]$
in~(\ref{eq:p_cov_general_form}), similar to~(\ref{eq:Pr_SINR_req_UAS1_LoS}), we have
\noindent
\begin{eqnarray}
 \textrm{Pr}\left[\frac{P\zeta_{n}^{\textrm{NL}}\left(r\right)h}{I_{r}+N_{0}}>\gamma\right]&\hspace{-0.4cm}=\hspace{-0.3cm}&\exp\hspace{-0.1cm}\left(\hspace{-0.1cm}-\frac{\gamma N_{0}}{P\zeta_{n}^{\textrm{NL}}\left(r\right)}\hspace{-0.1cm}\right)\hspace{-0.1cm}\mathscr{L}_{I_{r}}\hspace{-0.1cm}\left(\hspace{-0.1cm}\frac{\gamma}{P\zeta_{n}^{\textrm{NL}}\left(r\right)}\hspace{-0.1cm}\right).\hspace{0.5cm}\label{eq:Pr_SINR_req_wLT_UAS1_NLoS}
\end{eqnarray}


%

Our proof of Theorem~\ref{thm:p_cov_UAS2} is completed by plugging
(\ref{eq:geom_dis_PDF_UAS2_LoS}), (\ref{eq:geom_dis_PDF_UAS2_NLoS}), (\ref{eq:Pr_SINR_req_wLT_UAS1_LoS}), and (\ref{eq:Pr_SINR_req_wLT_UAS1_NLoS})
into (\ref{eq:p_cov_general_form}).


\section*{Appendix~B: Proof of Lemma~\ref{lem:laplace_term_UAS2_LoS_seg1}\label{sec:Appendix-B}}

Based on Theorem~\ref{thm:p_cov_UAS2}, it is straightforward
to derive $\mathscr{L}_{I_{r}}\left(s\right)$ in the range of $0<r\leq d_{1}$
as

\small
\noindent
\begin{eqnarray}
 \mathscr{L}_{I_{r}}\left(s\right)
 & \hspace{-0.3cm}=\hspace{-0.3cm} & \mathbb{E}_{\left[I_{r}\right]}\left\{ \left.\exp\left(-sI_{r}\right)\right|0<r\leq d_{1}\right\} \nonumber \\
 & \hspace{-0.3cm}=\hspace{-0.3cm} & \mathbb{E}_{\left[\Phi,\left\{ \beta_{i}\right\} ,\left\{ g_{i}\right\} \right]}\left\{ \left.\exp\left(-s\sum_{i\in\Phi/b_{o}}P\beta_{i}g_{i}\right)\right|0<r\leq d_{1}\right\} \nonumber \\
 & \hspace{-0.3cm}=\hspace{-0.3cm} & \mathbb{E}_{\left[\Phi\right]}\left\{ \left.\prod_{i\in\Phi/b_{o}}\mathbb{E}_{\left[\beta,g\right]}\left\{ \exp\left(-sP\beta g\right)\right\} \right|0<r\leq d_{1}\right\} \nonumber \\
 & \hspace{-0.3cm}\overset{(a)}{=}\hspace{-0.3cm} & \exp\left(\left.-2\pi\lambda\int_{r}^{\infty}\left(1-\mathbb{E}_{\left[g\right]}\left\{ \exp\left(-sP\beta\left(u\right)g\right)\right\} \right)udu\right|\right.\nonumber \\
 & \hspace{-0.3cm}\hspace{-0.3cm} & \left.\vphantom{\exp\left(\left.-2\pi\lambda\int_{r}^{\infty}\left(1-\mathbb{E}_{\left[g\right]}\left\{ \exp\left(-sP\beta\left(u\right)g\right)\right\} \right)udu\right|\right.}\qquad\qquad\qquad\qquad\qquad\qquad\qquad0<r\leq d_{1}\right),\label{eq:laplace_term_LoS_UAS1_seg1_proof_eq1}
\end{eqnarray}
\normalsize

\noindent where (a) in (\ref{eq:laplace_term_LoS_UAS1_seg1_proof_eq1})
is obtained from~\cite{Jeff's work 2011}.

Since $0<r\leq d_{1}$, $\mathbb{E}_{\left[g\right]}\left\{ \exp\left(-sP\beta\left(u\right)g\right)\right\} $
in (\ref{eq:laplace_term_LoS_UAS1_seg1_proof_eq1}) should consider
interference from both LoS and NLoS paths. Thus, $\mathscr{L}_{I_{r}}\left(s\right)$
can be further derived as

\small
\noindent
\begin{eqnarray}
\mathscr{L}_{I_{r}}\left(s\right) & \hspace{-0.3cm}= \hspace{-0.3cm}& \exp\left(-2\pi\lambda\int_{r}^{d_{1}}\left(1-\frac{u}{d_{1}}\right)\frac{u}{1+\left(sPA^{{\rm {L}}}\right)^{-1}u^{\alpha^{{\rm {L}}}}}du\right)\nonumber \\
 &  & \times\exp\left(-2\pi\lambda\int_{r_{1}}^{d_{1}}\frac{u}{d_{1}}\frac{u}{1+\left(sPA^{{\rm {NL}}}\right)^{-1}u^{\alpha^{{\rm {NL}}}}}du\right)\nonumber \\
 &  & \times\exp\left(-2\pi\lambda\int_{d_{1}}^{\infty}\frac{u}{1+\left(sPA^{{\rm {NL}}}\right)^{-1}u^{\alpha^{{\rm {NL}}}}}du\right).\label{eq:laplace_term_LoS_UAS1_seg1_proof_eq2}
\end{eqnarray}

\normalsize

\vspace{-0.1cm}
Plugging $s=\frac{\gamma r^{\alpha^{{\rm {L}}}}}{PA^{{\rm {L}}}}$ into~(\ref{eq:laplace_term_LoS_UAS1_seg1_proof_eq2}), and considering the definition of $\rho_{1}\left(\alpha,\beta,t,d\right)$ and $\rho_{2}\left(\alpha,\beta,t,d\right)$ in~(\ref{eq:rou1_func}) and~(\ref{eq:rou2_func}), we can obtain $\mathscr{L}_{I_{r}}\left(\frac{\gamma r^{\alpha^{{\rm {L}}}}}{PA^{{\rm {L}}}}\right)$ shown in~(\ref{eq:Lemma_6}), which concludes our proof.

\section*{\noindent Appendix C: Proof of Lemma~\ref{lem:laplace_term_UAS2_NLoS_seg1}\label{sec:Appendix-C}}

Following the same approach in Appendix~B, it is ready
to derive $\mathscr{L}_{I_{r}}\left(\frac{\gamma r^{\alpha^{{\rm {NL}}}}}{PA^{{\rm {NL}}}}\right)$
in the range of $0<r\leq x_{1}$ as

\small
\noindent
$\mathscr{L}_{I_{r}}\left(\frac{\gamma r^{\alpha^{{\rm {NL}}}}}{PA^{{\rm {NL}}}}\right) = $
\vspace{-0.1cm}
\begin{eqnarray}
 &&\exp\left(-2\pi\lambda\int_{r_{2}}^{d_{1}}\left(1-\frac{u}{d_{1}}\right)\frac{u}{1+\left(\frac{\gamma r^{\alpha^{{\rm {NL}}}}}{PA^{{\rm {NL}}}}PA^{{\rm {L}}}\right)^{-1}u^{\alpha^{{\rm {L}}}}}du\right)\nonumber \\
 &  & \times\exp\left(-2\pi\lambda\int_{r}^{d_{1}}\frac{u}{d_{1}}\frac{u}{1+\left(\frac{\gamma r^{\alpha^{{\rm {NL}}}}}{PA^{{\rm {NL}}}}PA^{{\rm {NL}}}\right)^{-1}u^{\alpha^{{\rm {NL}}}}}du\right)\nonumber \\
 &  & \times\exp\left(-2\pi\lambda\int_{d_{1}}^{\infty}\frac{u}{1+\left(\frac{\gamma r^{\alpha^{{\rm {NL}}}}}{PA^{{\rm {NL}}}}PA^{{\rm {NL}}}\right)^{-1}u^{\alpha^{{\rm {NL}}}}}du\right).\label{eq:proof_Lemma4_eq1}
\end{eqnarray}
\normalsize


Similarly, $\mathscr{L}_{I_{r}}\left(\frac{\gamma r^{\alpha^{{\rm {NL}}}}}{PA^{{\rm {NL}}}}\right)$
in the range of $x_{1}<r\leq d_{1}$ can be calculated
by

\vspace{0.2cm}
\small
\noindent
$\mathscr{L}_{I_{r}}\left(\frac{\gamma r^{\alpha^{{\rm {NL}}}}}{PA^{{\rm {NL}}}}\right) = $
\vspace{-0.1cm}
\begin{eqnarray}
&&\exp\left(-2\pi\lambda\int_{r}^{d_{1}}\frac{u}{d_{1}}\frac{u}{1+\left(\frac{\gamma r^{\alpha^{{\rm {NL}}}}}{PA^{{\rm {NL}}}}PA^{{\rm {NL}}}\right)^{-1}u^{\alpha^{{\rm {NL}}}}}du\right)\hspace{0.8cm}\nonumber \\
&&\times\exp\left(-2\pi\lambda\int_{d_{1}}^{\infty}\frac{u}{1+\left(\frac{\gamma r^{\alpha^{{\rm {NL}}}}}{PA^{{\rm {NL}}}}PA^{{\rm {NL}}}\right)^{-1}u^{\alpha^{{\rm {NL}}}}}du\right).\label{eq:proof_Lemma4_eq2}
\end{eqnarray}
\normalsize


Our proof is thus completed by plugging~(\ref{eq:rou1_func}) and~(\ref{eq:rou2_func}) into (\ref{eq:proof_Lemma4_eq1}) and
(\ref{eq:proof_Lemma4_eq2}).

\section*{Appendix D: Proof of Lemma~\ref{lem:laplace_term_UAS1_NLoS_seg2}\label{sec:Appendix-D}}

Following the same approach in Appendix~B,
it is ready to derive $\mathscr{L}_{I_{r}}\left(\frac{\gamma r^{\alpha^{{\rm {NL}}}}}{PA^{{\rm {NL}}}}\right)$
in the range of $r>d_{1}$ as

\small
\noindent
\begin{eqnarray}
\mathscr{L}_{I_{r}}\left(\frac{\gamma r^{\alpha^{{\rm {NL}}}}}{PA^{{\rm {NL}}}}\right) & \hspace{-0.3cm}=\hspace{-0.3cm} & \exp\left(-2\pi\lambda\int_{r}^{\infty}\frac{u}{1+\left(\gamma r^{\alpha^{{\rm {NL}}}}\right)^{-1}u^{\alpha^{{\rm {NL}}}}}du\right)\hspace{0.5cm}\nonumber \\
 & \hspace{-0.3cm}=\hspace{-0.3cm} & \exp\left(-2\pi\lambda\rho_{2}\left(\alpha^{{\rm {NL}}},1,\left(\gamma r^{\alpha^{{\rm {NL}}}}\right)^{-1},r\right)\right),\label{eq:proof_Lemma5_eq1}
\end{eqnarray}

\normalsize

\noindent where $\rho_{2}\left(\alpha,\beta,t,d\right)$ is defined
in (\ref{eq:rou2_func}).

Our proof is thus completed with (\ref{eq:proof_Lemma5_eq1}).

\end{comment}
\end{thebibliography}

\end{document}